\documentclass[graybox, nosecnum]{svmult}

\usepackage{mathptmx}       % selects Times Roman as basic font
\usepackage{helvet}         % selects Helvetica as sans-serif font
\usepackage{courier}        % selects Courier as typewriter font
\usepackage{type1cm}        % activate if the above 3 fonts are
\usepackage{makeidx}         % allows index generation
\usepackage{graphicx}        % standard LaTeX graphics tool
\usepackage{multicol}        % used for the two-column index
\usepackage[bottom]{footmisc}% places footnotes at page bottom
\usepackage{hyperref}        %for hyperlinks
\usepackage{soul}            % for high-lighting of text
\usepackage{xcolor}
\makeindex             % used for the subject index
\begin{document}

\title*{Emission of dispersive waves from solitons in axially-varying optical fibers}
\author{A. Kudlinski, A. Mussot, M. Conforti and D. V. Skryabin}
\institute{A. Kudlinski, A. Mussot, M. Conforti \at Univ. Lille, CNRS, UMR 8523 - PhLAM - Physique des Lasers Atomes et Mol\'ecules, F-59000 Lille, France, \email{alexandre.kudlinski@univ-lille1.fr}
\and D. V. Skryabin \at Department of Physics, University of Bath, Bath, United Kingdom \email{d.v.skryabin@bath.ac.uk}}
\maketitle
\abstract{The possibility of tailoring the guidance properties of optical fibers along the same direction as the evolution of the optical field allows to explore new directions in nonlinear fiber optics. The new degree of freedom offered by axially-varying optical fibers enables to revisit well-established nonlinear phenomena, and even to discover novel short pulse nonlinear dynamics. Here we study the impact of meter-scale longitudinal variations of group velocity dispersion on the propagation of bright solitons and on their associated dispersive waves. We show that the longitudinal tailoring of fiber properties allows to observe experimentally unique dispersive waves dynamics, such as the emission of cascaded, multiple or polychromatic dispersive waves.}

\section{Introduction}

Since its discovery in the frame of the Korteweg-de Vries equation by Zabusky and Kruskal \cite{Zabusky1965,Gardner1967}, the concept of solitons has been extended to many other systems described by integrable equations \cite{Ablowitz1973}, including the nonlinear Schr\"{o}dinger equation (NLSE) \cite{Zakharov1972} widely used to study nonlinear pulse propagation in optical fibers \cite{Mollenauer1980,Hasegawa2002,Dudley2006}. However, in real-world fibers, intrinsic higher-order dispersive and nonlinear effects break the integrability of the NLSE and therefore perturb the invariant propagation of solitons \cite{Kivshar1989}. Fundamental solitons, on the contrary to higher-order ones, are very stable in optical fibers and they usually survive these perturbations, by continuously adapting their temporal and spectral shapes \cite{Elgin1993}. The robustness of fundamental solitons has been exploited in the context of dispersion-managed optical communications in which the periodic evolution of losses and gain due to fiber attenuation and amplifiers is compensated by a periodic arrangement of dispersion and/or nonlinearity \cite{Turitsyn2012}. The so-called dispersion-managed solitons are propagating over kilometer-long systems and they are usually relatively broad (in the picosecond duration scale), making them weakly altered by higher-order dispersion. In fact, in some cases, third-order dispersion might even help to stabilize them \cite{Hizanidis1998}. In the case of much narrower solitons (in the hundreds of femtoseconds time scale), higher-order dispersion plays a much more significant role. In the case of a perturbation due to third-order dispersion for example, a short soliton propagating near the zero-dispersion wavelength (ZDW) loses energy into a dispersive wave (also called resonant or Cherenkov radiation) across the ZDW \cite{Wai1986,Wai1987} and experiences a spectral recoil in the opposite spectral direction in order to conserve the overall energy \cite{Akhmediev1995}. This very well-known process has been studied extensively from theoretical and experimental points of views \cite{Wai1986,Wai1987,Akhmediev1995,Cristiani2004,Erkintalo2012,Webb2013,Conforti2013}, in particular in the context of supercontinuum generation in which it plays a crucial role in the early dynamics \cite{Dudley2006,Skryabin2010}. Following their emission, dispersive waves can also collide with solitons in presence of Raman effect, leading to their nonlinear interaction \cite{Yulin2004,Efimov2005,Skryabin2005}. Understanding this nonlinear wave mixing process has been a key in extending supercontinuum sources towards the blue/ultraviolet spectral region \cite{Skryabin2010,Gorbach2006,Gorbach2007} and also towards long wavelengths in some specific cases \cite{Chapman2010}.

In this chapter, we study the process of dispersive wave emission from a soliton in various axially-varying optical fibers. We show that the longitudinal variation of guiding properties allows observe experimentally new and unique dynamics regarding dispersive waves. The first section introduces the basics of dispersive wave emission from a fundamental soliton and the second one focuses on the peculiarities of this process in axially-varying fibers.

\section{Emission of a dispersive wave from a soliton}\label{basics}

In this section, we will introduce the basic concepts of fundamental temporal soliton propagation in optical fiber with second-order dispersion [or group-velocity dispersion (GVD)] and Kerr nonlinearity, as well as radiation of dispersive wave in presence of third-order dispersion.

\subsection{Fundamental soliton}

The nonlinear propagation of light in dispersive and nonlinear optical fibers is described by the nonlinear Schr\"{o}dinger equation (NLSE)
\begin{equation}{\frac{\partial A}{\partial z} = -i\frac{\beta_2}{2} \frac{\partial^2 A}{\partial^2 t} + i\gamma \left|A\right|^2 A}
\label{nlse}
\end{equation}
Here, $t$ is the retarded time in the frame traveling at the group velocity $v_g$ of the input pulse. $A(z,t)$ is the envelope of the electric field, $\beta_2$ is the fiber GVD coefficient and $\gamma$ is the Kerr nonlinear parameter defined using the standard definition from \cite{Agrawal2012}. The fundamental soliton is an analytical solution of Eq. \ref{nlse} \cite{Zakharov1972} when the second-order dispersion coefficient is negative (anomalous GVD) and the nonlinear parameter $\gamma$ is positive. Its amplitude takes the form
\begin{equation}{A(z,t) = \sqrt{P_0}~\mathrm{sech}\left(\frac{t}{T_0}\right)}
\label{soliton}
\end{equation}
where $P_0$ is the soliton peak power and $T_0$ its duration. The soliton is termed \emph{fundamental} when the following condition is fulfilled
\begin{equation}{P_0 = \frac{\left| \beta_2 \right|}{\gamma T_0^2}}
\label{fund}
\end{equation}

\begin{figure}[b!]
\centering
\includegraphics[width=\linewidth]{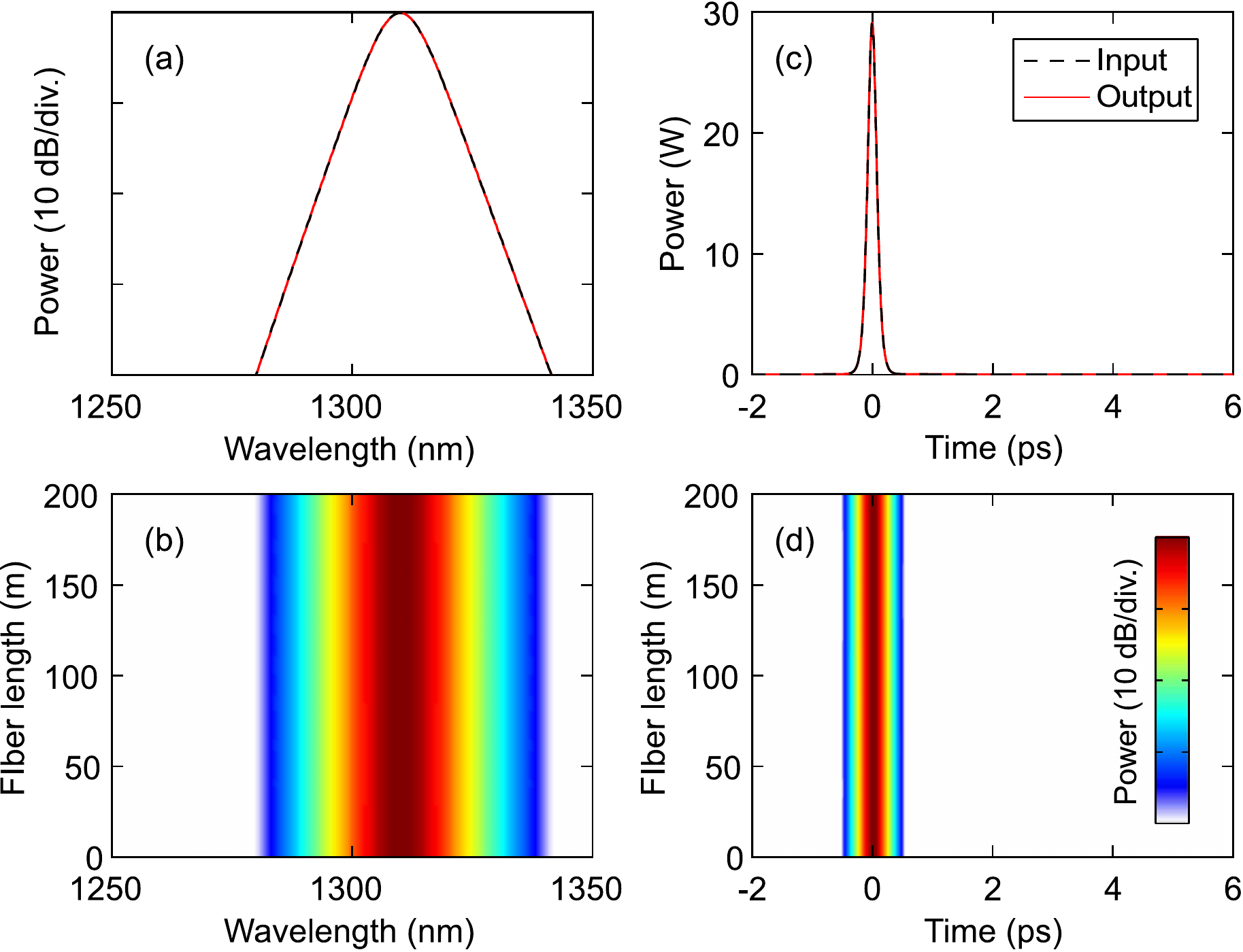}
\caption{Simulation of soliton propagation using Eq. \ref{nlse},\emph{ i.e.} neglecting third-order dispersion. (a) Input (black dashed line) and output (red solid line) spectra in a standard telecommunication single-mode fiber of 200 m length. (b) Spectral dynamics along propagation, with the same colorscale as in (d). (c) Corresponding input (black dashed line) and output (red solid line) temporal profiles. (d) Temporal dynamics along propagation. Simulation parameters are given in the text.}
\label{fig1}
\end{figure}

Figure \ref{fig1} shows the results of a numerical integration of Eq. \ref{nlse} in a standard telecommunication single-mode fiber. The second-order dispersion coefficient is $\beta_2 = -6.05 \times 10^{-28}$ s$^2$/m and the nonlinear parameter is $\gamma = 2.08$ W$^{-1}$.km$^{-1}$. The input pulse is a fundamental soliton centered at 1310 nm with a duration $T_0$ of 100 fs and a peak power of 29.08 W (in accordance with Eq. \ref{fund}).
Figures \ref{fig1}(a,b) and (c,d) show that the input pulse propagates without any spectral or temporal modification along the fiber, respectively. This is the most striking feature of temporal fundamental solitons and this is the main reason why they have attracted so much interest both from fundamental and applicative point of views \cite{Hasegawa2002,Taylor1996,Mollenauer1988}.

\subsection{Dispersive wave}

\begin{figure}[h!]
\centering
\includegraphics[width=\linewidth]{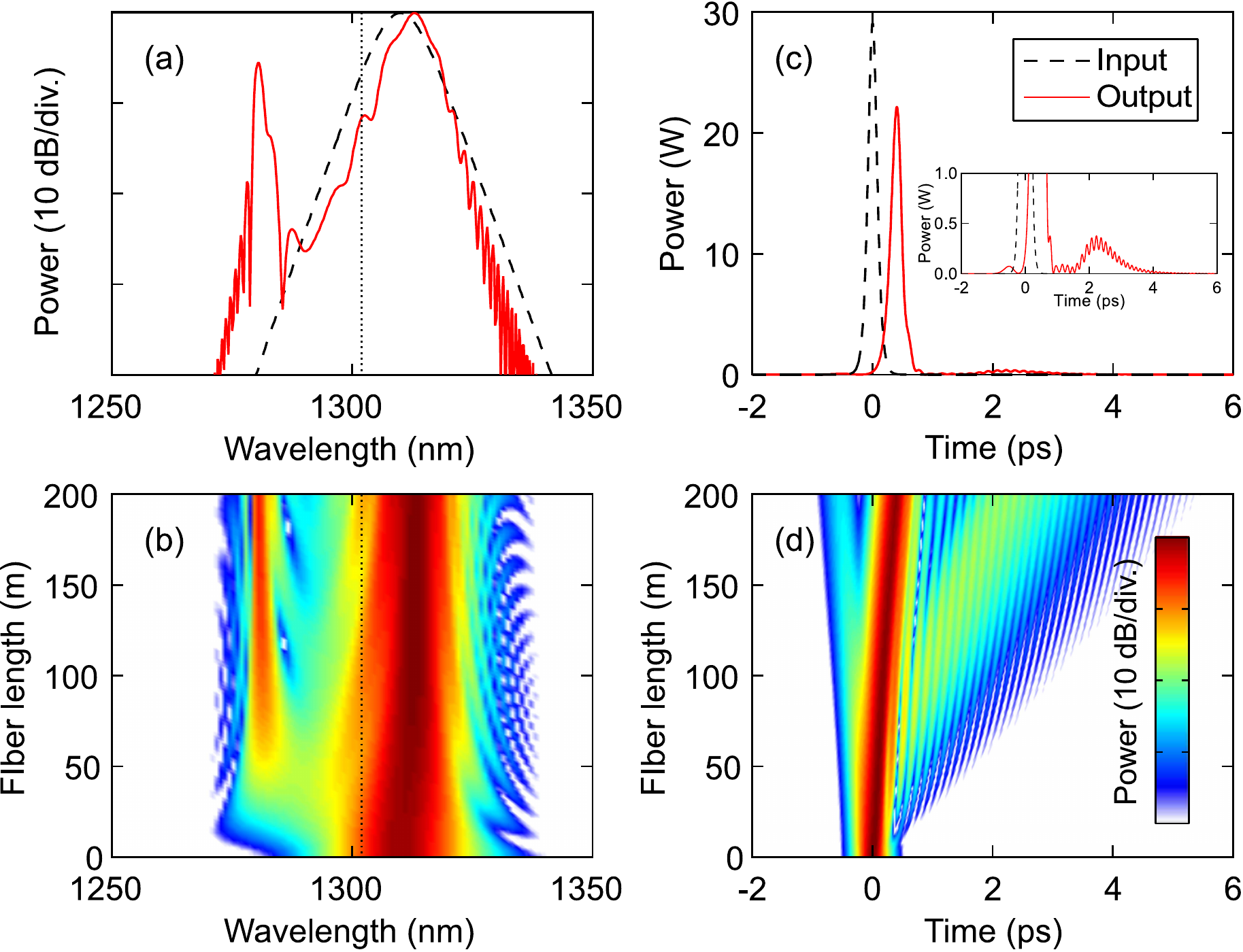}
\caption{Simulation of soliton propagation using Eq. \ref{nlse3}, \emph{i.e.} taking third-order dispersion into account. (a) Input (black dashed line) and output (red solid line) spectra in a standard telecommunication single-mode fiber of 200 m length. (b) Spectral dynamics along propagation, with the same colorscale as in (d). (c) Corresponding input (black dashed line) and output (red solid line) temporal profiles. (d) Temporal dynamics along propagation. Simulation parameters are given in the text. The black dotted lines in (a) and (b) represent the fiber ZDW. }
\label{fig2}
\end{figure}

In practice however, one cannot always neglect the fiber third-order dispersion. In particular, it plays a major role when the soliton is located near the zero-dispersion wavelength (ZDW) of the fiber \cite{Wai1986,Wai1987}, \emph{i.e.} when the soliton spectrum overlaps the normal GVD region located across the ZDW. In this case, Eq. \ref{nlse} has to be rewritten in order to take into account the fiber third-order dispersion. It takes the form
\begin{equation}
{\frac{\partial A}{\partial z} = -i\frac{\beta_2}{2} \frac{\partial^2 A}{\partial^2 t} +\frac{\beta_3}{6} \frac{\partial^3 A}{\partial^3 t} + i\gamma \left|A\right|^2 A}
\label{nlse3}
\end{equation}
where $\beta_3$ is the third-order dispersion coefficient of the fiber. The third-order dispersion drastically modifies the soliton dynamics as compared to the case in which it is neglected. This is illustrated in Fig. \ref{fig2} where we have simulated the propagation of the same input pulse as above using Eq. \ref{nlse3}. We have considered a $\beta_3$ value of $6.9 \times 10^{-41}$ s$^3$/m, the other fiber parameters being the same ones as above. The fiber ZDW [represented by the black dotted line in Figs \ref{fig2}(a) and (b)] is 1302 nm.
The output spectrum obtained after 200 m [red solid line in Fig. \ref{fig2}(a)] strongly differs from the input one (black dashed line). First, it exhibits a strong peak around 1280 nm corresponding to a dispersive wave. Second, the soliton spectrum is distorted and its central wavelength has red shifted from 1310 nm to 1313 nm. This is the spectral recoil accompanying the emission of the dispersive wave. The spectral dynamics displayed in Fig. \ref{fig2}(b) shows that the spectrum initially broadens towards the short-wavelength region and that the dispersive wave starts to form at about 50 m of propagation. In the time domain [Fig. \ref{fig2}(c)], it appears at the trailing edge of the soliton (red solid line) as a broad and relatively distorted pulse (see inset). Since the soliton has experienced a spectral recoil, it has slightly decelerated as compared to the fiber input. It has also lost some energy, which has been radiated into the dispersive wave. This can be further observed in the temporal dynamics plot of Fig. \ref{fig2}(d), where we can see the slow deceleration of the soliton as well as the continuous radiation of a broad dispersive wave along propagation.

The frequency at which the dispersive wave is emitted can be predicted from phase-matching arguments between the two radiations as \cite{Akhmediev1995}
\begin{equation}
\frac{\beta_2}{2}\Omega^2 + \frac{\beta_3}{6}\Omega^3 = \frac{\gamma P_0}{2}
\label{pm}
\end{equation}
where $\Omega = \omega_{S} - \omega_{DW}$ is the frequency separation between the soliton at $\omega_{S}$ and the dispersive wave at $\omega_{DW}$. Solving Eq. \ref{pm} with the parameters of our study gives a dispersive wave wavelength of 1283 nm, in excellent agreement with the numerical simulation results of Figs \ref{fig2}(a) and (b). The process of dispersive wave emission from a soliton is very well known and explained as long as optical fibers are uniform in length: when the soliton spectrum crosses the ZDW, the part of soliton energy located in the normal GVD region is radiated into a dispersive wave. This causes a spectral recoil of the soliton which thus moves away from the ZDW so that the emission process becomes less and less efficient and cannot occur again. Consequently, there is a unique dispersive wave which is generated and appears as a sharp spectral peak whose frequency does not change with propagation, as observed from Fig. \ref{fig2}(b). However, the process can be radically different in optical fibers which are not uniform as a function of length,\emph{ i.e.} in axially-varying fibers. This will be the focus of the remaining of this chapter.

\section{Generation of dispersive waves from solitons in axially-varying optical fibers}\label{generation}

\subsection{Axially-varying optical fibers}

For the fabrication of axially-varying optical fibers, the longitudinal evolution of the fiber diameter is controlled by adjusting the evolution of drawing speed with time (which is related to fiber length) using a servo-control system. This process is ruled by the conservation of glass mass between the preform and the fiber:
\begin{equation}\label{outerdiameter}
d_{\mathrm{Fiber}}=d_{\mathrm{Preform}}\sqrt{\frac{V_{\mathrm{Preform}}}{V_{\mathrm{Fiber}}}}
\end{equation}
where $d_{\mathrm{Preform, Fiber}}$ are respectively the preform and fiber outer diameter, and $V_{\mathrm{Preform, Fiber}}$ are respectively the preform feed into the furnace and the drawing capstan speed. In our process, $d_{\mathrm{Preform}}$ and $V_{\mathrm{Preform}}$ are fixed, $V_{\mathrm{Fiber}}$ is adjusted with a desired $f(z)$ function (where $z$ is the longitudinal space coordinate along the fiber), which results in a modulation of the fiber outer diameter $d_{\mathrm{Fiber}}(z)$, and thus of the overall fiber structure. This results in a modulation of the mode(s) propagation constant(s), and thus of all guiding properties.

\begin{figure}[h!]
\centering
\includegraphics[width=\linewidth]{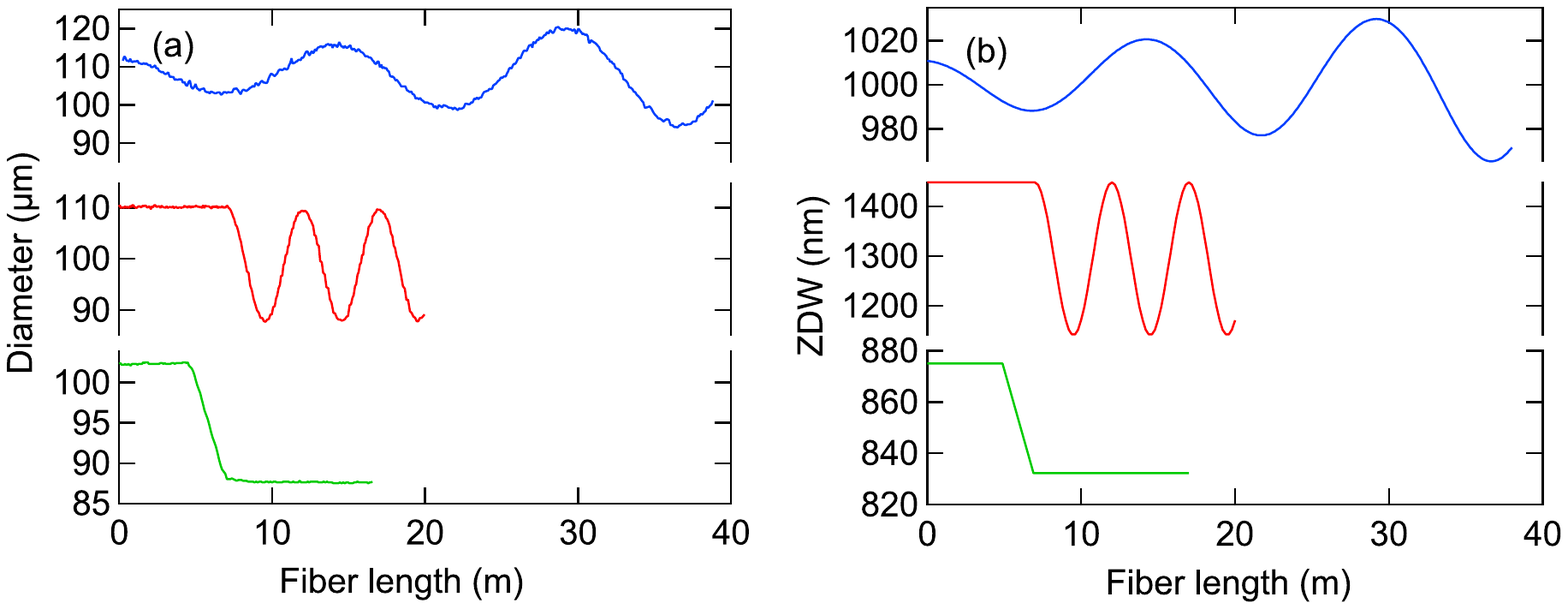}
\caption{(a) Evolution of the outer diameter as a function of fiber length for three different axially-varying fibers measured during the fiber fabrication process. (b) Corresponding calculated ZDW. For the second fiber (red case), the ZDW plotted in (b) corresponds to the second one (located at long wavelengths).}
\label{fiber}
\end{figure}

Figure \ref{fiber}(a) shows three examples of axially-varying fibers used hereafter. These curves show the evolution of the outer diameter recorded during the fiber drawing process. Figure \ref{fiber}(b) shows the corresponding calculated ZDW. The example represented in red corresponds to a fiber with two ZDWs, and the plot in Fig. \ref{fiber}(b) corresponds to the second ZDW. In the range of diameter variations investigated here, the ZDW approximately follows a linear dependance with fiber diameter.

\subsection{Emission of multiple dispersive waves along the fiber}

\begin{figure}[h!]
\centering
\includegraphics[width=\linewidth]{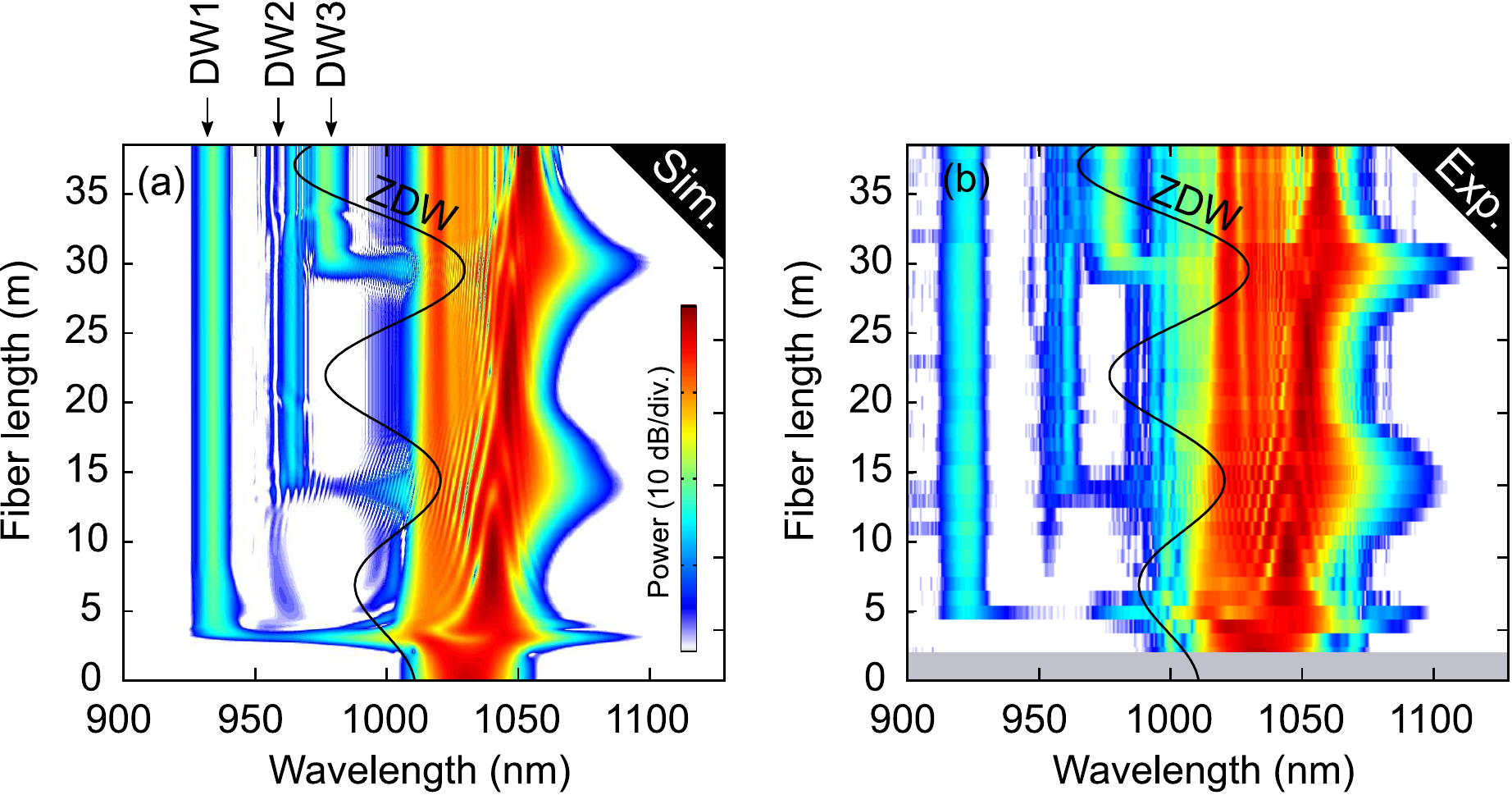}
\caption{(a) Numerical simulations and (b) experimental results in the spectral domain versus fiber length showing the emission of three distinct dispersive waves from a single soliton in an axially-varying fiber. The evolution of the ZDW with fiber length is represented by black solid lines. }
\label{fig3}
\end{figure}

As recalled above, in uniform fibers, the soliton experiences a red shift (spectral recoil) during the generation of a dispersive wave. This red shift is usually strongly reinforced by the Raman-induced soliton self-frequency shift. As a consequence, the soliton moves far away from the ZDW, so that the dispersive wave emission process cannot occur again. We will first show an example in which the use of a suitable axially-varying fiber allows for a single soliton to emit multiple dispersive waves along the fiber. Indeed, this can occur if the fiber ZDW moves along propagation so that it "hits" the soliton during its redshift following the emission of a dispersive wave \cite{Arteaga-Sierra2014,Billet2014}. In this case, the overlap between the soliton spectrum and the normal GVD region can become large enough to initiate the emission of a new dispersive wave. This is illustrated with numerical simulations and experiments in Figs. \ref{fig3}(a) and (b), respectively. For these numerical simulations, the stimulated Raman scattering term has been added to Eq. \ref{nlse3}, which now writes:
\begin{equation}
{\frac{\partial A}{\partial z} = -i\frac{\beta_2}{2} \frac{\partial^2 A}{\partial^2 t} +\frac{\beta_3}{6} \frac{\partial^3 A}{\partial^3 t} + i\gamma\bigg(A\int R(t')|A(t-t')|^2 dt' \bigg)}
\label{gnlse}
\end{equation}
where $R(t)=(1-f_R)\delta(t)+f_R h_R(t)$ includes both Kerr and Raman effects, where $h_R(t)$ correspond to the Raman response function ($f_R=0.18$) taken from \cite{Hollenbeck2002}.

We consider a fiber with a ZDW which evolves along the fiber [blue curve in Figs. \ref{fiber}(a) and (b)], following the profile represented by the black solid lines in Fig. \ref{fig3}(a) and (b). The ZDW is 1011 nm at the fiber input and reaches 1021 nm and 1030 nm at 14 m and 29 m, respectively. Simulations and experiments are performed with an input pulse of 410 fs full width at half maximum (FWHM) duration centered around 1030 nm and a peak power of 46 W. In order to be consistent with experiments, we consider a slight chirp with a chirp parameter of $+3.7$. All parameters can be found in Ref. \cite{Billet2014}. The simulation in Fig. \ref{fig3}(a) shows the emission of a first dispersive wave around 925 nm at a length of 5 m, in a similar fashion to what happens in uniform fibers (see Fig. \ref{fig2}). The soliton experiences a red shift due to the combined action of spectral recoil and Raman effect. The ZDW increases to 1021 nm at 14 m, which brings it closer to the soliton and therefore enhances the spectral overlap between the soliton and the normal GVD region. This initiates the emission of a second dispersive wave around 960 nm at this location in the fiber. The same process occurs a third time around 29 m, resulting in the emission of a third dispersive wave. Experiments reported in Fig. \ref{fig3}(b) are in excellent agreement with simulations and demonstrate the process of multiple dispersive wave emission from a single soliton.

\subsection{Cascading of dispersive waves}\label{cascade}

The process of dispersive wave emission can also occur in a slightly different scenario in uniform fibers: a soliton experiencing a strong Raman-induced self-frequency shift can hit the second ZDW of a fiber (located at long wavelengths), which cancels the red shift and generates a dispersive wave across the ZDW, at longer wavelengths \cite{Skryabin2003,Biancalana2004}. Here we will study this process in an axially-varying fiber [red curve in Figs. \ref{fiber}(a) and (b)] with two ZDWs separated by a region of anomalous dispersion in which a fundamental soliton is excited. The evolution of the second ZDW (the long wavelength one) is represented by the white solid lines in Figs. \ref{cascade}. It is constant over the first 7 m and then oscillates as a function of length. We consider input pulses of 340 fs FWHM duration centered around 1030 nm, with a chirp parameter of $+1.5$ and a peak power of 75 W. Figures \ref{cascade}(a) and (b) show respectively the numerical simulations performed with Eq. \ref{gnlse} and experimental results.

The input short pulse excites a fundamental soliton which initially experiences Raman-induced soliton self-frequency shift. Around 9 m, the ZDW has reached 1140 nm and is very close to the soliton so that a dispersive wave [labelled DW1 in Fig. \ref{cascade}(a)] is emitted across the ZDW, in the normal GVD region. At this point, the process is very similar to the one known in uniform fibers. The real novelty comes slightly after 10 m, when DW1 crosses the ZDW (which is increasing again at this point): as soon as DW1 crosses the ZDW, a new radiation (labelled CDW1) is generated at even longer wavelengths, around 1350 nm. A careful analysis of this process reveals that the continuously evolving GVD prevent the dispersive wave to strongly spread out in time [as expected in uniform fibers, see Fig. \ref{fig2}(d)] and allows to keep it relatively localized in time as a short pulse \cite{Bendahmane2014}. As a consequence, when crossing the ZDW, this pulse can emit another dispersive wave which we term cascaded dispersive wave (CDW1), in analogy with cascaded four-wave mixing processes.
\begin{figure}[t!]
\centering
\includegraphics[width=\linewidth]{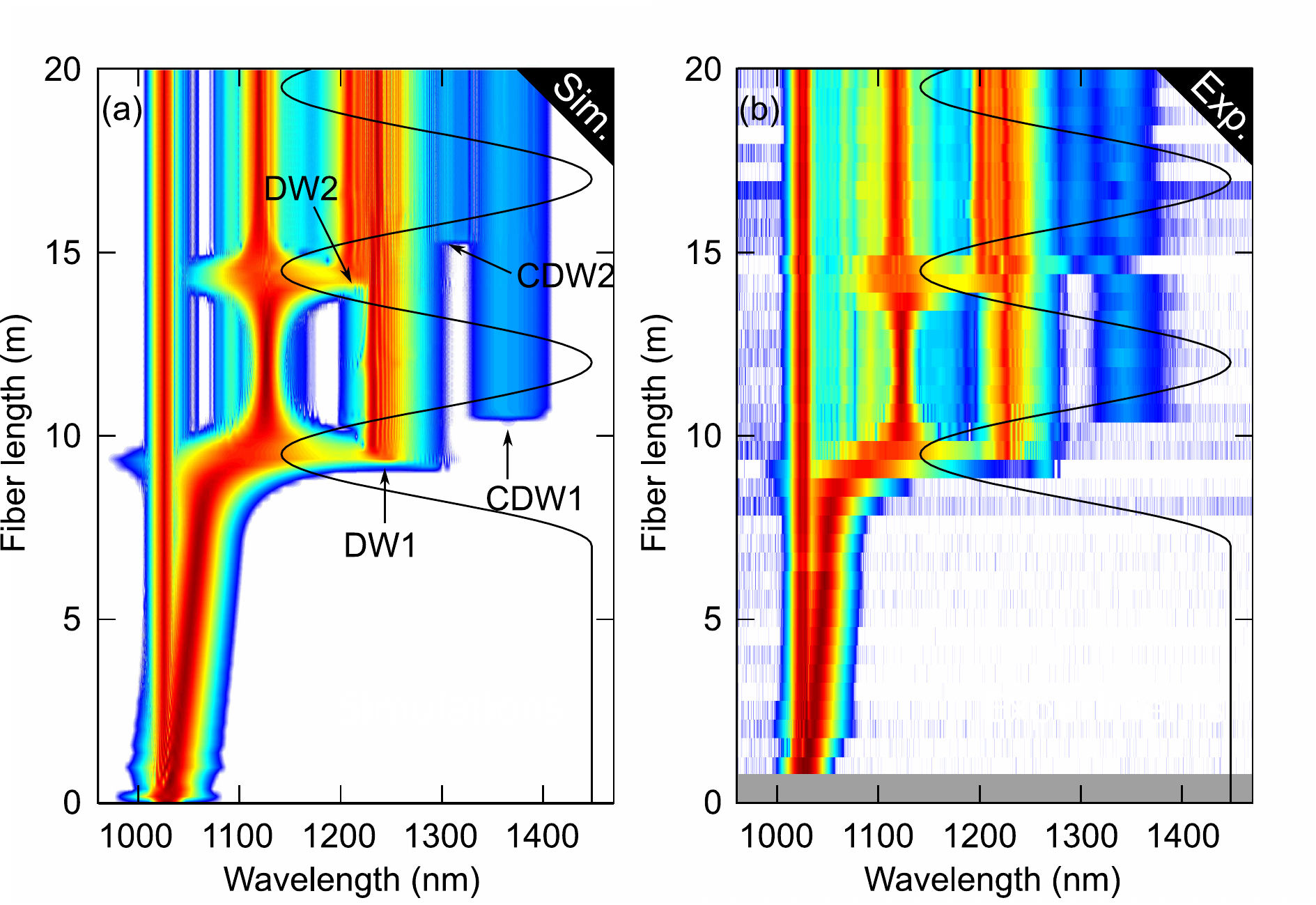}
\caption{(a) Numerical simulations and (b) experimental results in the spectral domain versus fiber length showing the emission of two cascaded dispersive waves in an axially-varying fiber with two ZDWs. The evolution of the second ZDW (located at long wavelengths) with fiber length is represented by black solid lines. }
\label{cascade}
\end{figure}

The exact same process occurs again farther in the fiber, around 15 m. The ZDW decreases again very close to the remaining of the soliton located slightly above 1100 nm so that a dispersive wave, DW2, is emitted. When DW2 crosses the ZDW slighty after, a cascaded dispersive wave, CDW2, is emitted. Experiments reported in Fig. \ref{cascade}(b) show again excellent agreement with numerical results, and provide the evidence for the process of cascaded dispersive wave generation. We might also note that, similarly to the previous section, multiple dispersive waves (DW1 and DW2) have been observed from a single soliton, around the second ZDW.

\subsection{Transformation of a dispersive wave into a fundamental soliton}

As mentioned above and deeply analyzed in Ref. \cite{Bendahmane2014}, the cascaded dispersive wave process is due to the fact that the dispersive wave initially generated experiences a GVD that varies with length, which allows it to remain temporally localized as a pulse to initiate the generation of a cascaded dispersive wave. Here, we will study this process more into details. For that, we consider a soliton launched in the vicinity of the ZDW so that it emits a dispersive wave in the normal GVD region, similarly to the usual case illustrated in Fig. \ref{fig2}. Once the dispersive wave is emitted, the fiber parameters are changed so that the GVD at the dispersive wave wavelength becomes anomalous [green curve in Figs. \ref{fiber}(a) and (b)] and the evolution of the radiation is studied.
\begin{figure}[b!]
\centering
\includegraphics[width=\linewidth]{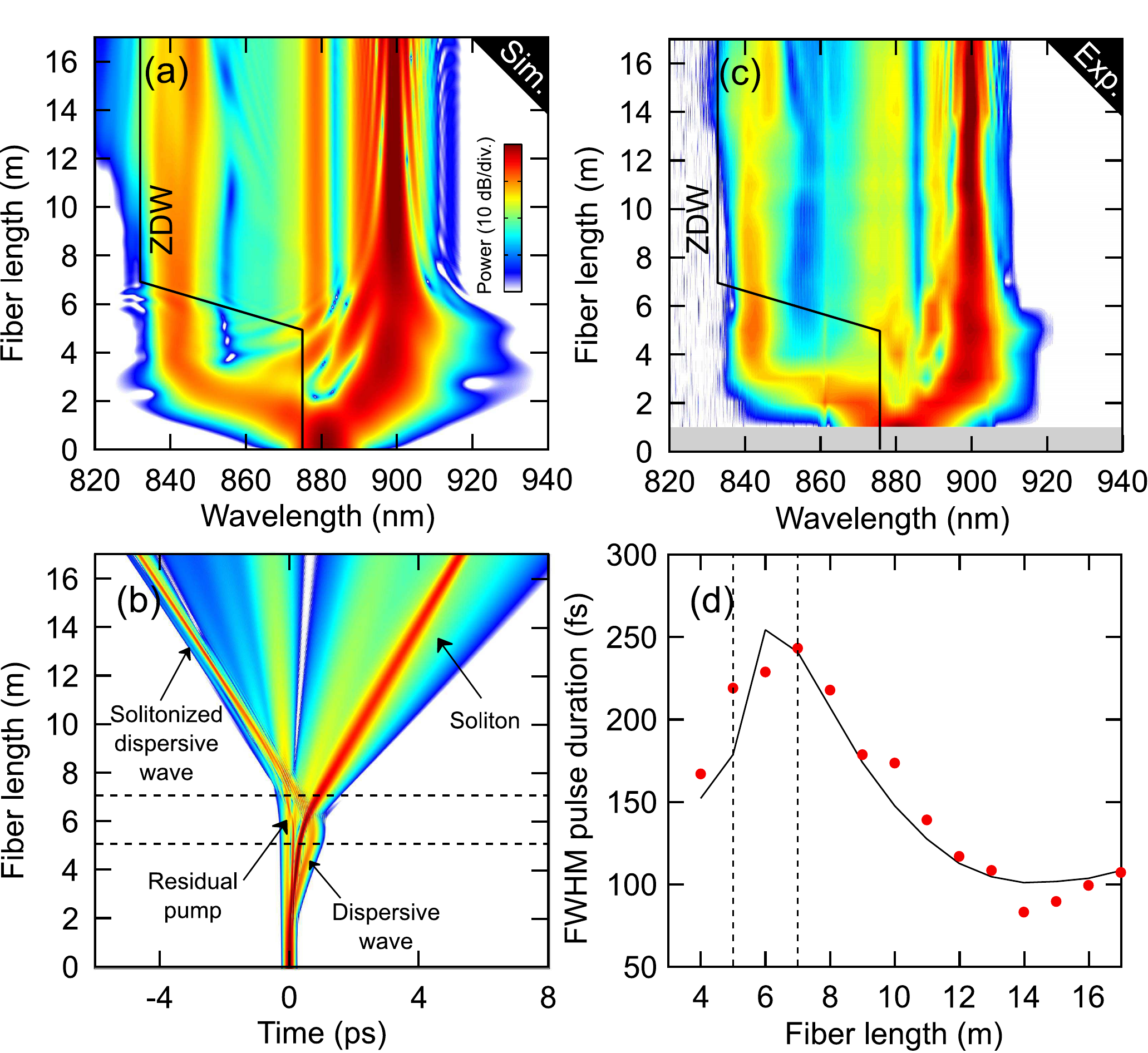}
\caption{(a,b) Numerical simulations and (c,d) experimental results showing the transformation of a dispersive wave into a fundamental soliton. (a) and (b) correspond respectively to the simulated spectral and temporal evolutions. (c) Measured spectral evolution versus fiber length. (d) Measured pulse duration around 840 nm (red full circles) and simulated one (black solid line). Black solid lines in (a) and (c) depict the ZDW. Dashed lines in (b) and (d) depict the limits of the tapered section. }
\label{solitonization}
\end{figure}
The results are summarized in Fig. \ref{solitonization}. The input pulse is transform limited with a 140 fs FWHM duration. It is centered around 881 nm and has a peak power of 42 W. The spectral evolution versus fiber length [simulations in Fig. \ref{solitonization}(a) and experiments in Fig. \ref{solitonization}(c)] shows the emission of a dispersive wave around 840 nm but does not highlight much change after the GVD change occurring between 5 and 7 m. The simulated time domain simulation of Fig. \ref{solitonization}(b) provides much more information about the dynamics. The soliton decelerates due to the combined effects of spectral recoil and Raman-induced self-frequency shift and the dispersive wave, which starts spreading out, is located slightly behind it, as expected. Once it enters the tapered region (materialized by the two dashed lines), the dispersive wave accelerates due to changing dispersion and crosses the soliton. At the same time, it recompresses and further propagated as a short pulse in the remaining of the fiber, recalling the behavior of a soliton. In fact, further theoretical analysis reveals that the dispersive wave has indeed been transformed into a fundamental soliton \cite{Braud2016} and propagates as such in the remaining anomalous dispersion region. Experimental autocorrelation measurements reported in Fig. \ref{solitonization}(d) confirm this behavior: the pulse duration initially increases and then decreases after the tapered region, where they are perfectly fitted by square hyperbolic secant functions \cite{Braud2016}. These results show that a dispersive wave emitted from a soliton can be itself transformed into a fundamental soliton by carefully varying the longitudinal evolution of dispersion.

\subsection{Emission of polychromatic dispersive waves}\label{poly}

In the previous section, we have shown that a dispersive pulse initially travelling in normal GVD can become transformed into a fundamental soliton when entering a region with anomalous dispersion. Here we will study a reversed situation in which a fundamental soliton propagating in anomalous dispersion enters a region with normal dispersion. Fundamentally, the pulse cannot be a soliton anymore, and it is expected to linearly disperse. We will see that it can excite a so-called polychromatic dispersive wave \cite{Milian2012,Kudlinski2015}.

\begin{figure}[h!]
\centering
\includegraphics[width=\linewidth]{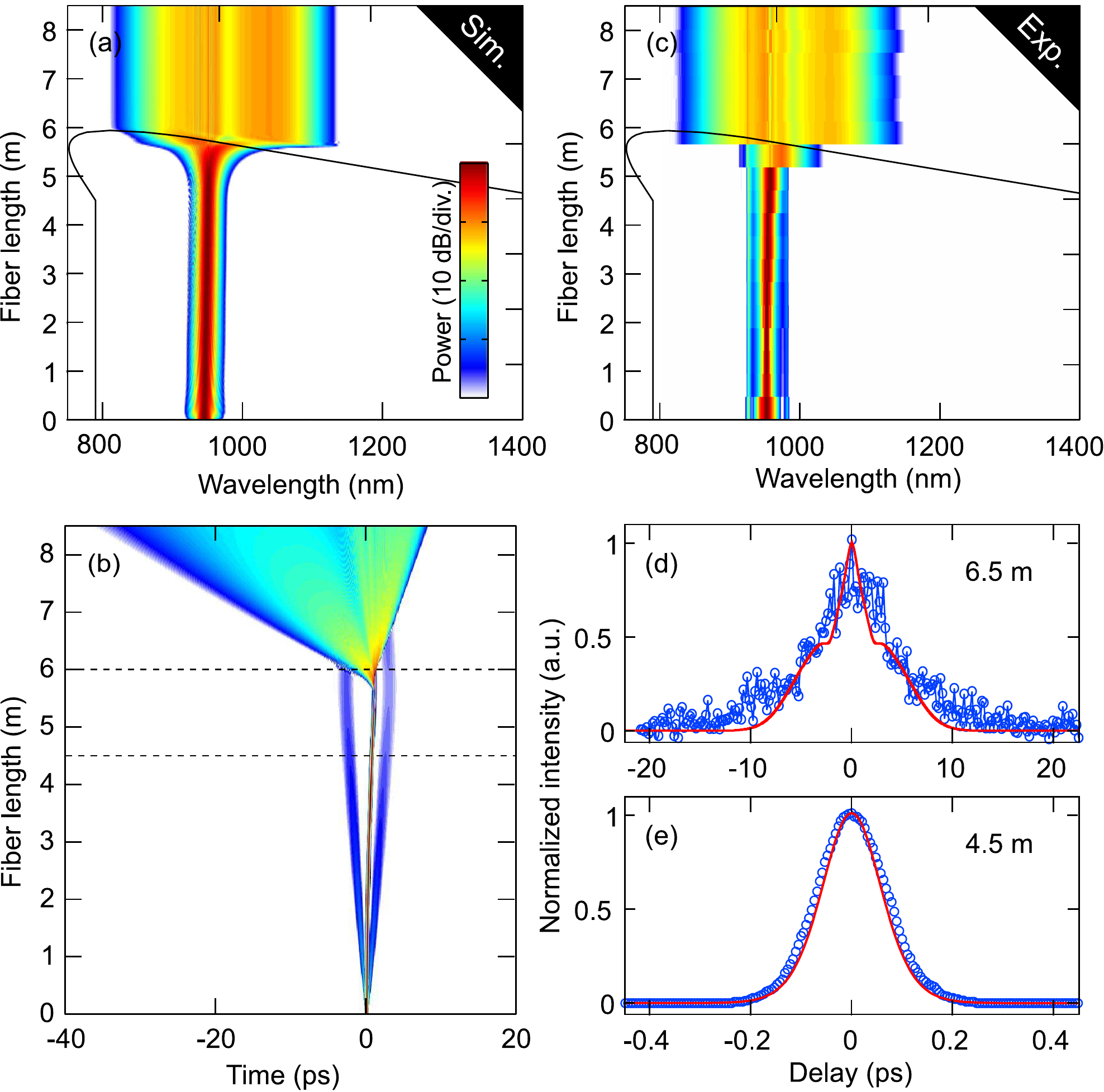}
\caption{(a,b) Numerical simulations and (c,d,e) experimental results showing the annihilation of a fundamental soliton into a polychromatic dispersive wave. (a) and (b) correspond respectively to the simulated spectral and temporal evolutions. (c) Measured spectral evolution versus fiber length. (d,e) Measured autocorrelation trace (open blue circles) and simulated ones (red solid line) at fiber length of (e) 4.5 m and (d) 6.5 m. Black solid lines in (a) and (c) depict the ZDWs. Dashed lines in (b) depict the limits of the tapered section. }
\label{polychromatic}
\end{figure}

Here, a fundamental soliton is excited by launching a transform limited pulse with a 130 fs FWHM duration around 950 nm, with a peak power of 110 W. The fiber is uniform over the first 4.5 m and then it is tapered down so that both ZDWs [depicted by solid lines in Figs. \ref{polychromatic}(a) and (c)] decreases until they join each other at 6 m. After this point, the fiber has all normal GVD all over the spectral range of interest here. The spectral dynamics, displayed in Figs. \ref{polychromatic}(a) and (c) for simulations and experiments respectively, show the initial propagation of a soliton until 5.5 m, where it crosses the second ZDW and thus enters the normal GVD region. a this point, a spectacular spectral broadening occurs, indicating that the pulse experiences a nonlinear. In fact, as long as the soliton spectrum starts to cross the ZDW, the emission of a dispersive wave is initiated. Since the second ZDW keeps decreasing and crossing the soliton, there is a continuous emission of radiation into the dispersive wave. Because the GVD properties of the fiber continuously changes, the phase-matching relation \ref{pm} linking the soliton and the dispersive wave continuously changes too, resulting in the generation of a broad radiation termed polychromatic dispersive wave \cite{Milian2012,Kudlinski2015}.

The process is as spectacular in the time domain plot of Fig. \ref{polychromatic}(b), where it can be seen that the significant broadening occurring at around 6 m is accompanied by a strong temporal broadening, which is consistent with the generation of a dispersive, strongly chirped pulse. This was confirmed experimentally by recording autocorrelation traces before and after the tapered section. At 4.5 m [Fig. \ref{polychromatic}(e)], the pulse has a duration of 113 fs (blue markers) and is well fitted by a square hyperbolic secant function, in good agreement with the simulation result (red solid line), which confirms its solitonic nature. But soon after the tapered section, at 6.5 m, the autocorrelation trace is much longer and has a much distorted profile, again in agreement with simulations. At this point, there is no clue of the presence of a soliton, which has therefore totally been annihilated into a polychromatic dispersive wave.

\subsection{Generation of a dispersive wave continuum}

In this section, we will investigate a complex scenario in which multiple solitons and dispersive waves are generated in an axially-varying fiber. This results in the generation of a 500 nm supercontinuum exclusively composed of dispersive waves.

For that, we use the same configuration as in section \ref{cascade} except that we increase the peak power of the input pulse to 380 W. This exceed the required power to form a fundamental soliton given by Eq. \ref{fund}. In this case, the input pulse excites a higher-order soliton which immediately breaks up into several fundamental solitons \cite{Beaud1987}.
\begin{figure}[b!]
\centering
\includegraphics[width=\linewidth]{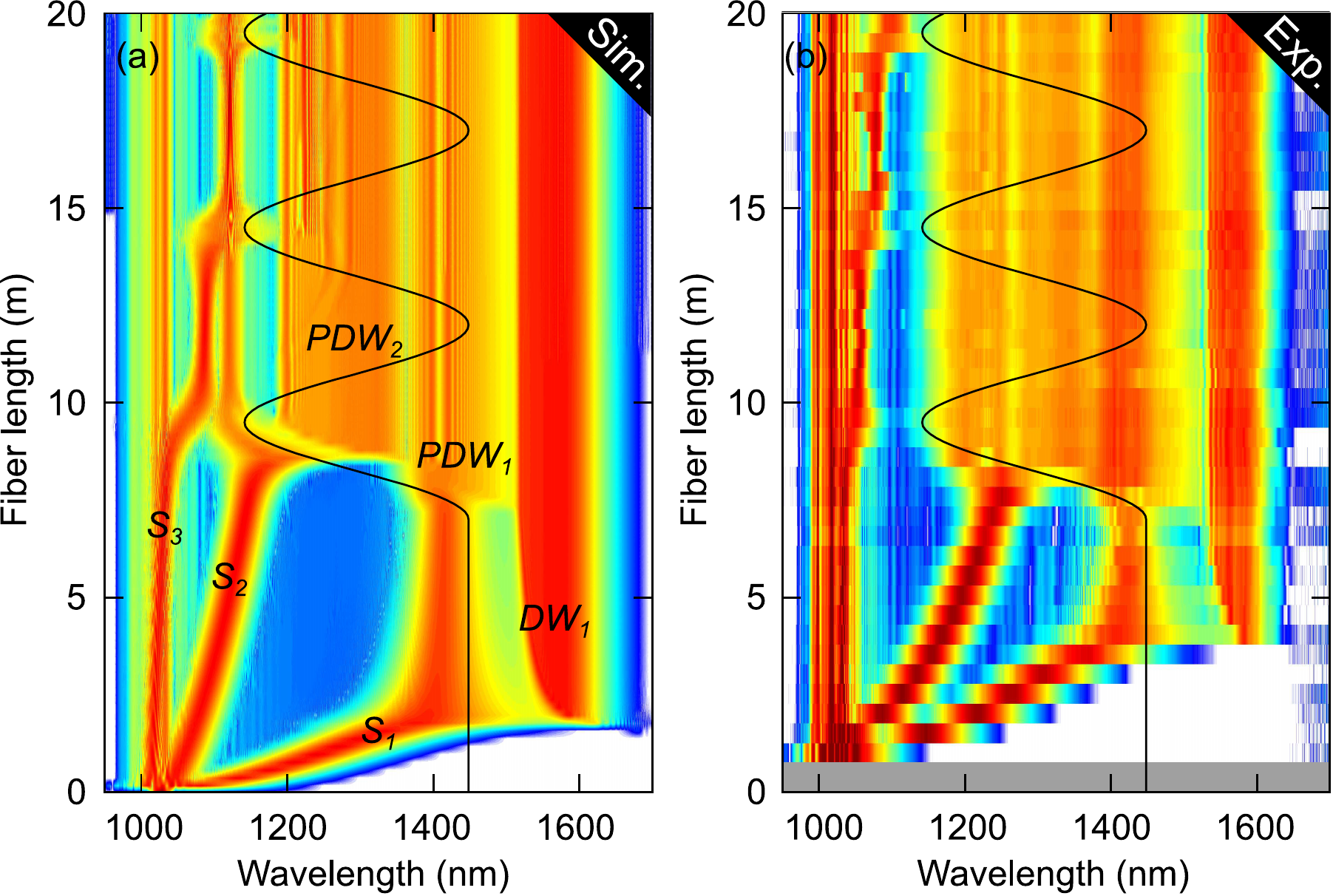}
\caption{(a) Numerical simulations and (b) experimental results in the spectral domain versus fiber length showing the generation of a dispersive wave continuum in an axially-varying fiber with two ZDWs. The evolution of the second ZDW (located at long wavelengths) with fiber length is represented by black solid lines. }
\label{sc}
\end{figure}Figures \ref{sc}(a) and (b) show respectively the simulated and experimental spectral dynamics recorded in this case. Three main solitons, labelled $S_1$ to $S_3$, form here. The first one, $S_1$, is also the most powerful one \cite{Lucek1992} and therefore experiences the most efficient Raman-induced frequency shift. It rapidly hits the second ZDW (in black solid line) so that the frequency shift stops and a strong dispersive wave (labelled $DW_1$) is emitted around 1600 nm. The soliton is frequency-locked around 1400 nm and hits the ZDW which starts to decrease at around 7 m. At this point, it emits a polychromatic dispersive wave (labelled $PDW_1$) following the process described in section \ref{poly}. The second soliton ($S_2$) experiences a less significant Raman-induced self-frequency shift than $S_1$ and hits the ZDW around 8 m, where it is already decreasing. Therefore, it emits a broad polychromatic dispersive wave (labelled $PDW_2$). Soliton $S_3$ remains relatively far from the ZDW all over the propagation and thus does not contribute significantly to the emission of dispersive waves.

Finally, the output spectrum between 1200 nm and 1600 nm is exclusively composed of dispersive waves and polychromatic dispersive waves generated from the two most powerful solitons.

\section{Conclusion and perspectives}

The perturbation of a fundamental soliton by third-order dispersion in optical fibers causes the emission of a resonant dispersive wave across the zero dispersion point. In axially-varying fibers, the guiding properties can be tailored as a function of propagation distance so that the ZDW continuously evolves as the soliton propagates. This allows to observe specific dispersive waves dynamics, such as the emission of cascaded, multiple or polychromatic dispersive waves.

Such axially-varying fibers can also be used to control the properties of the soliton itself, such as harnessing its Raman-induced redshift \cite{Judge2009,Bendahmane2013} or even induce a blueshift \cite{Stark2011}. In periodic fibers, multiple quasi phase matched dispersive waves can also be observed thanks to the periodicity \cite{Conforti2015,Conforti2016}, following a completely different mechanism than the one presented here \cite{Conforti2015}. More generally, this work illustrates to remarkable robustness of fundamental solitons against various types of perturbations, and the extremely rich nonlinear dynamics that these perturbations can induce.

\section*{Acknowledgements}

We acknowledge Beno\^{\i}t Barviau, Abdelkrim Bendahmane, Maximilien Billet, Flavie Braud, Andy Cassez, Tomy Marest, Stefano Trillo and Shaofei Wang.

\bibliographystyle{osa}
%\bibliographystyle{osa}
%\bibliography{biblio_solitons}
%\section{Other Options}
%\runinhead{Run-in Heading Boldface Version} Use the \LaTeX\ automatism for all your citations.
%
%\subruninhead{Run-in Heading Italic Version} Use the \LaTeX\ automatism for all your citations.

%\input{referenc}
\end{document}